\documentclass[11pt,a4paper,english]{article}
\usepackage{color}
\usepackage{amsmath}    
\usepackage{amssymb}
\usepackage{graphicx} 
\usepackage{float}
\usepackage{authblk}
\usepackage[a4paper]{geometry}
\usepackage[final]{pdfpages}

\usepackage{soul} 

\begin{document}     

\title{Jensen's force and the statistical mechanics of
  cortical asynchronous states}

\author[1,2,3]{Victor Buend{\'\i}a}
\author[1]{Pablo Villegas}
\author[4]{Serena di Santo} 
\author[2,5]{Alessandro Vezzani} 
\author[2,3]{Raffaella Burioni}
\author[1,2]{Miguel A. Mu\~noz}
  
  \affil[1]{Departamento de Electromagnetismo y F{\'\i}sica de la
  Materia e Instituto Carlos I de F{\'\i}sica Te\'orica y
  Computacional. Universidad de Granada.  E-18071, Granada, Spain}
  \affil[2]{Dipartimento di Matematica, Fisica e Informatica,
  Universit\`a di Parma, via G.P. Usberti, 7/A - 43124, Parma, Italy}
  \affil[3]{INFN, Gruppo Collegato di Parma, via G.P. Usberti, 7/A -
  43124, Parma, Italy}
  \affil[4]{Scuola Internazionale Superiore
  di Studi Avanzati, via Bonomea, 265 - 34136 Trieste, Italy.}
  \affil[5]{IMEM-CNR, Parco Area delle Scienze 37/A - 43124 Parma, Italy}

\date{}  
\maketitle  
  
\begin{abstract}

  The cortex exhibits self-sustained highly-irregular activity even
  under resting conditions, whose origin and function need to be fully
  understood. It is believed that this can be described as an
  "asynchronous state" stemming from the balance between excitation
  and inhibition, with important consequences for
  information-processing, though a competing hypothesis claims it
  stems from critical dynamics.  By analyzing a parsimonious
  neural-network model with excitatory and inhibitory interactions, we
  elucidate a noise-induced mechanism called "Jensen's force"
  responsible for the emergence of a novel phase of arbitrarily-low
  but self-sustained activity, which reproduces all the experimental
  features of asynchronous states.  The simplicity of our framework
  allows for a deep understanding of asynchronous states from a broad
  statistical-mechanics perspective and of the phase transitions to
  other standard phases it exhibits, opening the door to reconcile,
  asynchronous-state and critical-state hypotheses.  We argue that
  Jensen's forces are measurable experimentally and might be relevant
  in contexts beyond neuroscience.
\end{abstract}


Networks of excitatory units --in which some form of ``activity''
propagates between connected nodes-- are successfully used as abstract
representations of propagation phenomena as varied as epidemics,
computer viruses, or memes in social networks \cite{Romu-Review}. Such
dynamical processes can be either in an \emph{active} phase in which
activity reverberates indefinitely through the network or in a
\emph{quiescent} phase where activity eventually ceases; in some cases
of interest they lie at the very edge of the quiescent/active phase
transition \cite{Henkel,Marro,RMP}.

Some systems of outmost biological relevance cannot be, however,
modeled as networks of purely excitatory units.  Nodes that inhibit
(or repress) further activations are essential components of
neuronal circuits in the cortex \cite{wilson1972}, as well as of
gene-regulatory, signaling, and metabolic networks
\cite{GRN,metabolic}.  Indeed, an essential feature of cortical
networks is that they are composed of both excitatory and inhibitory
neurons; synaptic excitation occurs always in concomitance with
synaptic inhibition.  What is the function of such a co-occurrence of
excitation and inhibition? or, quoting a recent review article on the
subject, ``why should the cortex simultaneously push on the
accelerator and on the brake?''  \cite{Scanziani}.

Generally speaking, inhibition entails much richer sets of dynamical
patterns including oscillations \cite{toggles,Chen} and
counterintuitive phenomena. For example, in a nice and intriguing
paper that triggered our curiosity, it was argued that inhibition
induces ``ceaseless'' activity in excitatory/inhibitory (E/I) networks
\cite{Larremore}.  More in general, inhibition helps solving a
fundamental problem in neuroscience, namely, that of the dynamic
range, defined as follows. Each neuron in the cortex is connected to
many others, but individual synapses are relatively weak, so that each
single neuron needs to integrate inputs from many others to become
active; this leads to an explosive, all-or-none type of recruitment in
populations of purely excitatory neurons, i.e. to a discontinuous
phase transition between a quiescent and an active phase
\cite{Scanziani}.  In other words, the network is either quiescent or
almost saturated. This would severely constrain the set of possible
network states, hindering the network capacity to produce diverse
responses to differing inputs.  This picture changes dramatically in
the cortex, where the presence of inhibition has been empirically
observed to allow for much larger dynamic ranges owing to a
progressive (smoother) recruitment of neuronal populations
\cite{Liu,Pouille}.  This is consistent with the well-known empirical
fact that neurons in the cerebral cortex remain slightly active even
in the absence of stimuli \cite{Softky,Arieli,Abeles}.  In such a
state of low self-sustained activity neurons fire in a steady but
highly-irregular fashion at a very low rate and with little
correlations among them.  This is the so-called \emph{asynchronous
  state}, which has been argued to play an essential role for diverse
computational tasks \cite{Yuste,Vogels,recent-Sompo,Machens}.

It has become widely accepted that such an asynchronous
state of low spontaneous activity emerges from the interplay between
excitation and inhibition.  Models of \emph{balanced} E/I networks, in
which excitatory and inhibitory inputs largely compensate each other,
constitute --as it was first theoretically proposed
\cite{vanW-S1,vanW-S2,Brunel1,Brunel2,Renart} and then experimentally
confirmed \cite{McCormick,McCormick2,Trevino,Destexhe2016,Reyes}-- the
basis to rationalize asynchronous states. Indeed, balanced E/I
networks are nowadays considered as a sort of ``standard model'' of
cortical dynamics \cite{Latham2015}.

In spite of solid theoretical and experimental advances, a full
understanding of the phases of E/I networks remains elusive. For
instance, it is still not clear if simple mathematical models can sustain
highly-irregular low-activity phases even in the complete absence of
external inputs from other brain regions. Indeed, many existing
approaches to the asynchronous state assume that it requires of
external inputs from other brain regions to be maintained
\cite{Destexhe2009}, while some others rely on endogenously firing
neurons --i.e. firing even without inputs-- for the same purpose (see
e.g. \cite{Latham2000}).  Furthermore, it is not clear from modelling
approaches whether asynchronous states can have very low (rather than
high or moderate) levels of activity \cite{Destexhe2009,Einevoll,brasileiros}.

All these problems can be summarized --from a broader Statistical
Mechanics perspective-- saying that it is not well-understood whether
the asynchronous state constitutes an actual physical phase of
self-sustained activity different from the standard quiescent and
active ones. It is not clear either if novel non-standard types of
phase transitions emerge at its boundaries.  Such possible phase
transitions might have important consequences for shedding light in to
the so-called ``criticality hypothesis''. This states that the cortex
might operate close to the edge of a phase transition to optimize its
performance; thus, it is essential to first understand what the
possible phases and phase transitions are.  

Here, we analyze the simplest possible network model including
excitation and inhibition in an attempt to create a parsimonious model
--understood as the simplest possible yet not-trivial model-- of E/I
networks \cite{Larremore}.  We show, by employing a combination of
theoretical and computational analyses, that the introduction of
inhibitory interactions into purely excitatory networks leads to a
self-sustained low-activity phase intermediate between conventional
quiescent and active phases.  Remarkably, the novel phase stems from a
noise-induced mechanism that we call ``Jensen's force''  (or ``Jensen's drift'')
--for its relationship with Jensen's inequality in probability
theory-- and that occurs owing to the combined effect of inhibition
and network sparsity.  The low-activity intermediate phase shares all
its fundamental properties with asynchronous states and thus, as we
argue, our model constitutes the simplest possible
statistical-mechanics representation of asynchronous endogenous
cortical activity.  Moreover, continuous (critical) phase transions
--separating the novel intermediate phase from the quiescent and
active phases, respectively-- are elucidated, with possible important
consequences to shed light on the criticality hypothesis
\cite{Mora-Bialek,RMP,LG}, and to make an attempt to reconcile the
asynchronous-state and criticality hypotheses, putting them together
within a unified framework. Finally, we propose that the elucidated
Jensen's force might be relevant in other contexts such as
e.g. gene regulatory networks.

\section{Models and Results} 
\subsection{Minimal model} 
The simplest approach to capture the basic elements of E/I networks
are two-state (binary) neuron models \cite{vanW-S1,recent-Sompo}, such
as the one proposed by Larremore {\emph et al.} \cite{Larremore}.  The
simplified version that we consider here consists of a random-regular
directed network with $N$ nodes and $K$ links \cite{Newman-Review}. A
fraction $\alpha$ of the nodes (typically $\alpha=0.2$ to mimic
empirical observations in the cortex \cite{Soriano,Reyes}) are
\emph{inhibitory} (negative interactions) and the rest are excitatory
(positive interactions). More specifically, the network is
\emph{hyper-regular}, meaning that not only all nodes have the same
inbound and outbound connectivity $k=K/N$, but also that each of them
receives exactly $\alpha k$ inhibitory inbound links and
$(1-\alpha) k$ of excitatory ones (see Fig.1 and Methods).

At any given (discrete) time $t$ the state of a single node, $i$, can
be either active, $s_i(t)=1$, or inactive $s_i(t)=0$.  The dynamics is
such that each node $i$ integrates the (weighted) activity of its $k$
neighbors as sketched in Fig.1. At time $t+1$, $s_i$ becomes active
(resp. inactive) with probability $\mathcal{P}_i$
(resp. $1-\mathcal{P}_i$) given by
\begin{equation}
\begin{array}{c}
  \mathcal{P}_i\equiv f
  \left(\Lambda_i={\displaystyle\frac{\gamma}{ k}}
  {\sum_{j}}\omega_{ij}s_{j}\left(t\right)\right)=\begin{cases}
  0 & \Lambda_i<0\\
  \Lambda_i & 0\leq\Lambda_i\leq1\\
  1 & \Lambda_i>1
\end{cases}
\end{array}
\label{model}
\end{equation}
where $f$ is a transfer function of the input $\Lambda_i$, $j$ runs
over the set of ($k$) nodes pointing to node $i$, $w_{ij}$ is the
weight of the connection from node $j$ to node $i$
($\omega_{ij}=\pm1$, for simplicity), and $\gamma$ is the overall
coupling-strength that acts as a control parameter.

The model is kept purposely simple in an attempt to reveal the basic
mechanisms of its collective behavior; more complex network
architectures, transfer functions, and other realistic ingredients are
implemented \emph{a posteriori} to verify the robustness of the
results.
\begin{figure}[h]
\centering
\includegraphics[width=0.8\columnwidth]{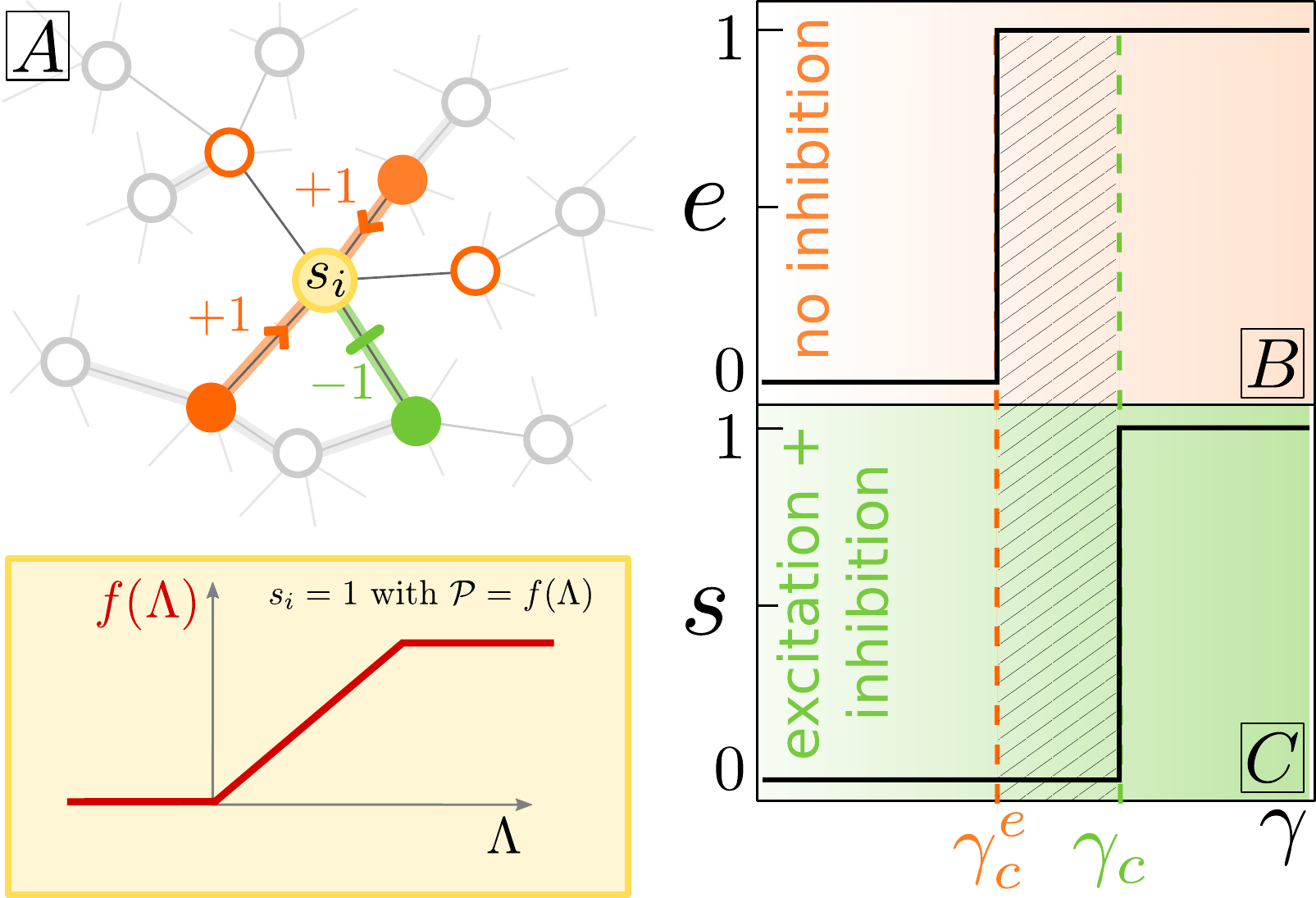}
\caption{(A) Upper panel: Sketch of the input received by a single
  node, including excitatory (orange arrows) and inhibitory (green
  blunt arrows) interactions from active (colored) neighbors.  The
  lower panel shows the considered transfer function for probabilistic
  activation of nodes as a function of the input.  (B) Averaged level
  of activity in a fully-connected network consisting solely of
  $N(1-\alpha)$ excitatory nodes; it exhibits a discontinuous phase
  transition at $\gamma_{c}^{e}(\alpha)=1/(1-\alpha)$ separating a
  quiescent or Down state from an active or Up one. (C) As (B) but for
  a network consisting of $N(1-\alpha)$ excitatory and $N\alpha$
  inhibitory nodes. Let us remark that the shape of the phase
  transition depends on our choice for the transfer function. More
  plausible, non-linear, transfer functions lead e.g. to discontinuous
  transitions with a region of bistability (phase coexistence) and
  hysteresis; however, the main results of this work are remain
  unaffected (see Supplementary Information (SI) 2).  }\label{MField}
\end{figure}

\subsection{Mean-field approach: massively connected networks: } 
We start considering the case of a fully connected network.  Let $E$
and $I$ be the total number of excitatory and inhibitory active nodes,
respectively, at a given time. These evolve stochastically according
to a Master equation (as described in Methods), from which
--performing a $1/N$ expansion-- one readily obtains the following
deterministic equations:
$\dot{e} = \left(1-\alpha\right)\langle f\left(\Lambda
\right)\rangle-e$ and
$\dot{i} = \alpha\langle f \left( \Lambda \right)\rangle-i$ --where
the dot stands for time derivative- for $e=E/N$ and $i=I/N$,
respectively. It follows that, in the steady state, excitation and
inhibition are proportional to each other: $e/(1-\alpha)=i/\alpha$,
i.e. they become spontaneously balanced in a dynamical way. Moreover,
the overall activity density, $s=e+i$, obeys
\begin{equation}
\dot {s} = \langle f\left( \Lambda \right) \rangle - s,
\label{exactS}
\end{equation}
while the difference $q=e-i$ is simply proportional to $s$ in the
stationary state: $q= (1-2\alpha) s$.  In the large network-size limit
(i.e.  $ N \rightarrow \infty$), fluctuations in the input of each
node are negligible.  In such a limit, all nodes receive the same
input, and thus the mean-field approach, in which the mean of the
transfer function values (outputs) is replaced by the transfer
function of the mean input
\begin{equation}
\dot s = f\left(\langle \Lambda \rangle\right) -s ,
\label{MFS}
\end{equation}
becomes exact.  Eq.(\ref{MFS}) admits two trivial fixed points
corresponding to the quiescent ($s^*=0$) and saturated ($s^*=1$)
states, respectively. The quiescent (resp. saturated) state is stable
below a given value of the coupling constant,
$\gamma < \gamma_c=1/(1-2\alpha)$ (resp.  $\gamma> \gamma_c$), while
right at $\gamma_c$ all values of $0 \le s \le 1$ are marginally
stable. Thus, as illustrated in Fig.\ref{MField}B, the system
experiences a discontinuous phase transition at $\gamma_c$ (i.e. the
all-or-none phenomenon described in the Introduction).  Observe also
(see Fig.1C) that, in agreement with intuition, as the fraction of
inhibitory nodes in the network is increased (i.e. as $\alpha$ grows),
the overall level of activity tends to decrease, and the nature of the
phase transition is not altered: it remains discontinuous even in the
presence of inhibitory populations.

\subsection{Beyond mean-field: Sparse networks}
\begin{figure}[h]
\centering
\includegraphics[width=0.8\columnwidth]{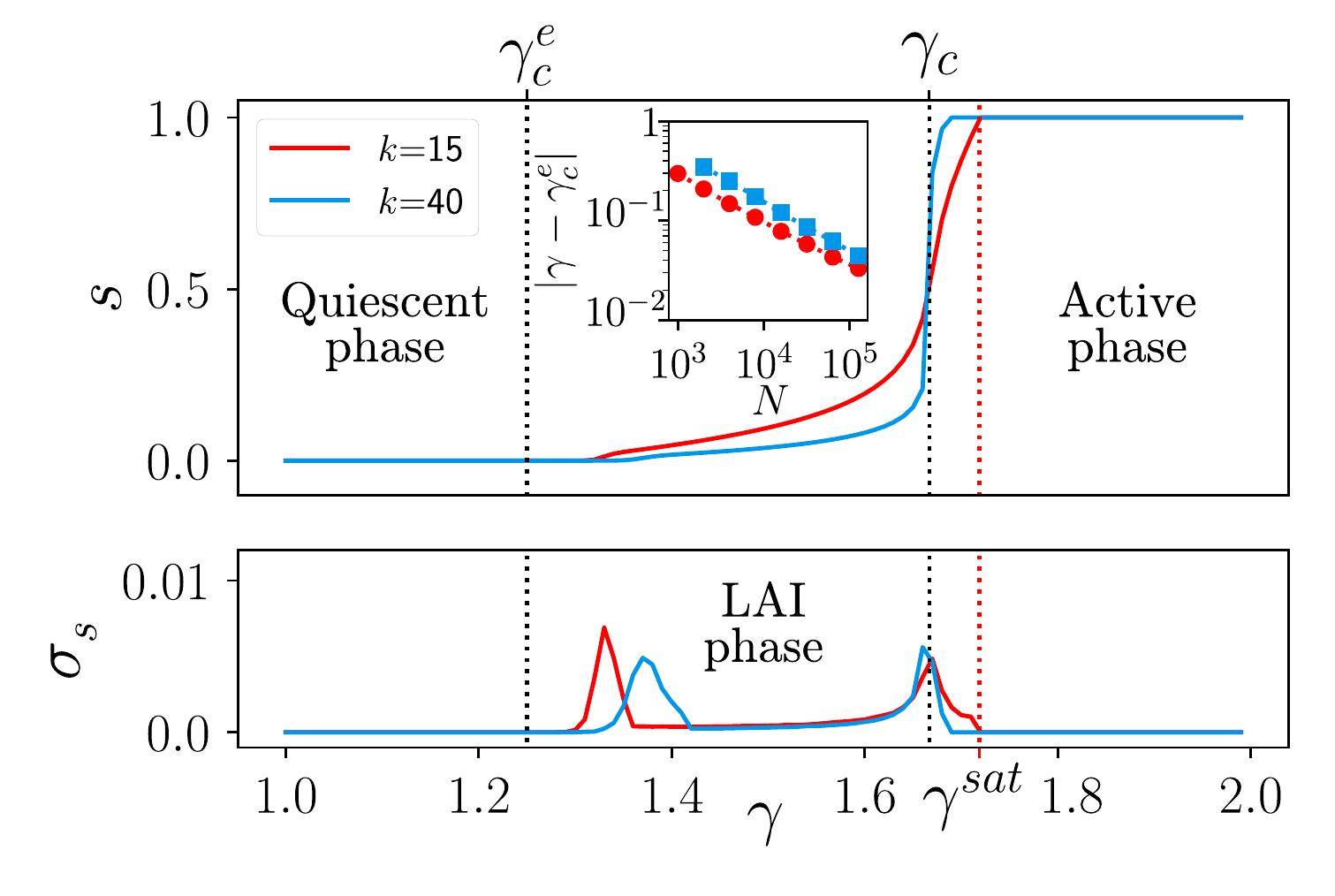}
\caption{Overall steady-state averaged network activity $s$
  for the E/I model on a sparse hyper-regular network ($N=16000$) in
  which all nodes have the same (in-)connectivity $k$ (with either
  $k=15$ or $k=40$) and the same fraction of ($(1-\alpha) k$)
  excitatory and ($\alpha k$) inhibitory inputs ($\alpha=0.2$ here).
  (A, Bottom) Variance across ($10^3$) runs of the total network
  activity averaged in time windows of a given length ($T=10^4$
  MonteCarlo steps) as a function of the coupling strength $\gamma$
  for two different values of the connectivity $k$; each curve shows
  two marked peaks, indicative of two phase transitions.  The leftmost
  one, $\gamma_c^e(k,N)$, shifts towards $\gamma_c^e$ in the large-$N$
  limit, obeying finite-size scaling, as illustrated by the straight
  line in the double-logarithmic plot of the inset.  On the other
  hand, the second peak is a remanent of the mean-field first-order
  transition at $\gamma_c=1/(1-2\alpha)= 1.66...$ and is hardly
  sensitive to finite-size effects.}.\label{PDiagram}
\end{figure}

Computational analyses of the model on sparse networks reveal a
phenomenology much richer than the just described mean-field one.  As
shown in Fig.\ref{PDiagram} the phase transition becomes progressively
smoother (continuous) as the network connectivity $k$ is reduced, and
a novel intermediate phase where the overall average activity $s$ does
not saturate to either $0$ or $1$ emerges.
Importantly, let us stress that such an intermediate phase does not
appear in sparse networks of purely excitatory nodes.

To gauge the level of network-state variability, we measured
computationally the standard deviation $\sigma_s$ of $\bar s$ (average
of $s$ finite-time windows for finite-size networks; see
Fig.\ref{PDiagram}) over realizations. This quantity exhibits two
marked peaks (Fig.\ref{PDiagram}) suggesting the existence of two
phase transitions \cite{Binney,Henkel}.  The (leftmost) peak, at
$\gamma_c ^e$, corresponds to a transition from the quiescent ($s=0$)
to the \emph{low-activity intermediate (LAI)} phase. Observe that,
$\gamma_c ^e$ exhibits severe finite-size-scaling corrections
(depending also on $k$) converging to $\gamma_{c}^{e}=1/(1-\alpha)$ as
$N \rightarrow \infty$ (see the inset in Fig.\ref{PDiagram}). This
value of $\gamma$ coincides with the mean-field transition point for
the purely excitatory subnetwork with $N(1-\alpha)$ units
(i.e. without inhibition; see Fig.1A), justifying the superindex $e$
in $\gamma_{c}^{e}$.  On the other hand, the second peak is located at
$\gamma_c =1/(1-2\alpha)$, i.e. the very same location of the
mean-field discontinuity for the fully-connected network.  These two
transition points delimit the LAI phase.  There is a third relevant
value, $\gamma=\gamma^{sat}$ (within the active phase) at which the
fully-saturated solution, $s=1$, emerges. As $k$ increases, this third
point becomes closer to $\gamma_c$, making the second transition
progressively sharper and converging to the mean-field result.

\subsection{Analytical results for sparse networks}
To rationalize the novel (LAI)phase with low levels of activity, it is
essential to realize that, in the sparse-connectivity case, the input
received by a given node does not necessarily take its mean-field
value, but is a fluctuating variable, making it thus necessary to
consider Eq.(\ref{exactS}) rather than its mean-field counterpart
Eq.(\ref{MFS}). To make analytical progress it is necessary to
determine the input distribution, which is equivalent to computing the
probability $p_{lj}(s)$ that a given node has exactly $l$ active
inhibitory neighbors and $j$ active excitatory ones, for arbitrary
values of $l$ and $j$.

Larremore {\emph{et. al.}} made an attempt to solve this problem
working with the actual (``quenched'') network architecture, which
requires scrutinizing the (spectral) properties of the associated
connectivity matrices \cite{Larremore}. Here, we propose to tackle the
problem from a complementary angle. More specifically, we consider a
random-neighbor (``annealed'') network version of the model, in which,
at each time step, the neighbors of each node are randomly sampled
from the whole network (keeping fixed the number of them as well as
the fractions of excitatory and inhibitory ones).  This annealed
variant of the model greatly simplifies the analytical calculations,
and --quite surprisingly-- leads to results identical (up to numerical
precision) to those for the original quenched-network problem.

For the annealed version of the model one can readily write (see SI-3):
\begin{equation}
  p_{lj}(s)=\left(\begin{array}{c}
                                        k\alpha\\
                                        l
\end{array}\right)\left(\begin{array}{c}
k\left(1-\alpha\right)\\ j
\end{array}\right)s^{j+l}\left(1-s\right)^{k-j-l},
\label{plj}
\end{equation}
which depends solely on $s$, i.e. the probability for any arbitrary
node to be active.  From this, it follows that
\begin{equation}
 \langle f(\Lambda) \rangle= \underset{l,j}{\sum}p_{lj}(s)f[\tilde\gamma(j-l)],
\label{average}
\end{equation}
(where $\tilde\gamma=\gamma/k$), as well as
$\langle \Lambda \rangle= \gamma (1-2\alpha) s$ and
$\sigma^2(\Lambda) =  \gamma ^2 s(1-s)/k$ for the mean and the
variance of the input distribution, respectively.  Note that all these
are functions of $s$ and $\tilde\gamma$, solely.  Evaluating
Eq.(\ref{average}) is not straightforward owing to the non-linearity
of $f$. However, analytical insight can be obtained by Taylor-expanding
around either of the two trivial solutions: $s^*=0$ or
$s^*=1$. Expanding around $s^*=0$ and keeping only leading order
(linear in $s$) contributions, leads to
$\left\langle f\left(\Lambda\right)\right\rangle \simeq f\left(
  \tilde\gamma k\left(1-\alpha\right) s\right) , $ which plugged into
Eq.(\ref{exactS}) implies that
the solution $s^*=0$ loses its stability at a critical point
$\gamma_{c}^{e}=1/\left(1-\alpha\right)$, in perfect agreement with
the computational observations (for $N\rightarrow \infty$)
Observe that the LAI noise-induced phase exists for all finite
  connectivity values and emerges at $\gamma_{c}^{e}$ for all $k$, but
  --owing to finite-size corrections-- larger and large networks are
  required to see it as the network connectivity is increased.

  A similar analysis around $s^*=1$ (see SI-4) reveals that the
  saturated solution is stable only above
\begin{equation}
 \gamma^{sat}=\frac{1-k(1-\alpha)}{(1-\alpha)-k(1-\alpha)(1-2\alpha)},
\label{1}
\end{equation}
again in perfect agreement with numerical findings (see Fig.2 and
SI-4).  As expected $ \gamma^{sat}$ converges to the mean-field
prediction $1/(1-2\alpha)$ for $k\rightarrow \infty$, as numerically
observed.

Thus, contrarily to mean-field expectations, there exists a whole
intermediate region, $\gamma_c^e<\gamma < \gamma^{sat}$, where
activity does not vanish nor saturate for E/I networks. Such a region
emerges as a consequence of input fluctuations and, hence, stems from
network sparsity. Observe that for purely excitatory networks,
i.e. with $\alpha=0$, $\gamma_c^e = \gamma^{sat}$ and the 
intermediate region vanishes.
\begin{figure}[h]
\centering
\includegraphics[width=0.8\columnwidth]{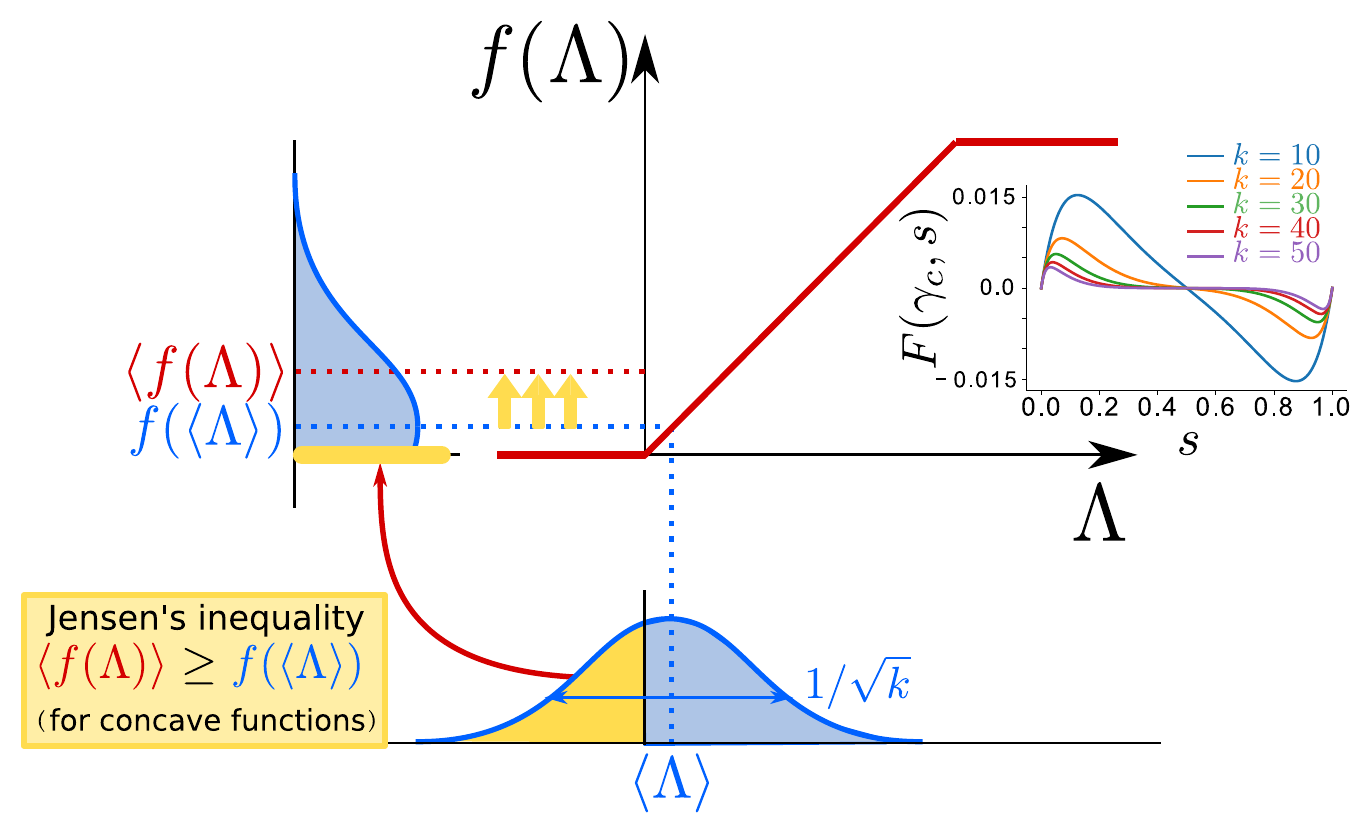}
\caption{Sketch illustrating the origin of the noise-induced Jensen's
   force. Each node in a sparse network receives an input
  $\Lambda$ which is a random variable extracted from some bell-shaped
  probability distribution function $P(\Lambda)$ (sketched below the
  x-axis) with averaged value
  $\langle \Lambda \rangle= \gamma (1-2\alpha) s$ and standard
  deviation $\sigma_s=(\gamma\sqrt{s(1-s)})/\sqrt{k} $ (see SI-3).
  The possible outputs $f(\Lambda)$ are also distributed according to
  some probability (sketched to the left of the y-axis).  Given that
  around $\Lambda\approx 0$ the function $f(\Lambda)$ is locally
  convex then, as a consequence of Jensen's inequality for convex
  functions,
  $ \langle f(\Lambda) \rangle \ge f(\langle \Lambda \rangle)$
  (i.e. the dotted red line is above the blue one). Indeed, while for
  positive inputs, the transformation is linear, negative ones are
  mapped into $0$ thus creating a net positive Jensen's 
  force for small values of $\Lambda$ (or $s$). The inset shows the
  Jensen's  force
  $F(\tilde{\gamma},s) \equiv \langle f(\Lambda) \rangle> - f(\langle
  \Lambda \rangle)$ computed right at the critical point $\gamma_c$
  for different connectivity values, as a function of $s$. Note, the
  negative values for large values of $s$ which stem from the
  concavity of the function $f(x)$ around $x=1$.  Note that $F$
  decreases as $k$ grows and vanishes in the mean-field limit. }
\label{Jensen}
\end{figure}
\subsection{ Jensen's  force}
To go beyond perturbative results, note that the difference between
the exact equation for the model on a sparse network,
Eqs.(\ref{exactS}), and its mean-field approximation, Eqs.(\ref{MFS}),
is that
$\langle f(\Lambda)\rangle\neq f\left(\langle\Lambda\rangle\right)$,
i.e. the non-linear function $f$ and the network average are
non-commuting ``operators'', and the reported non-trivial effects for
sparse networks necessarily stem from the difference between them:
\begin{equation}
 F(\tilde{\gamma},s) \equiv \langle f(\Lambda)\rangle 
- f\left(\langle\Lambda\rangle\right). 
\label{jensen}
\end{equation}
Observe that, as the terms in the r.h.s. depend on $s$,
$F(\tilde{\gamma},s)$ is state-dependent stochastic force.  As shown
above (and as suggested by the central limit theorem) the distribution
of inputs to any given node is centered at $\langle \Lambda \rangle$
and has a standard deviation that scales as $1/\sqrt{k}$.  If $f$ was
a linear function, then
$\langle f(\Lambda) \rangle =f(\langle \Lambda \rangle) $, but as it
is a convex function near the origin, then the Jensen's inequality of
probability theory (which expresses the fact that the if $x$ is a
random variable and $g(x)$ is a convex function, then
$<g(x)> \ge g(<x>)$) implies that
$\langle f(\Lambda) \rangle > f(\langle \Lambda \rangle) $, i.e. $F$
is positive if $\langle \Lambda \rangle$ is near the $0$, i.e. if $s$
is relatively small.

Thus, we propose the term ``\emph{Jensen's force}'' to refer to
$F(\tilde{\gamma},s)$ (see Fig.\ref{Jensen}).  This positive force is
responsible for the destabilization of the quiescent state and the
emergence of the LAI phase.  Observe that if, on the other hand,
$\langle \Lambda \rangle$ happens to be close to $1$, the function $f$
is locally concave and, using a reverse argument,
$\langle f(\Lambda) \rangle < f(\langle \Lambda \rangle) $, i.e. there
is a negative Jensen's force $F$ in the regime of very large
activities (justifying the reduction of the saturated regime with
respect to the mean-field case).  Finally, if parameters are such that
the system lies in the quiescent ($s=0$) or in the saturated ($s=1$)
phase then there are no input fluctuations --i.e. the input
distribution is delta function-- and the Jensen's force vanishes.

$F(\tilde{\gamma},s)$ can be analytically calculated for some
particular transfer functions $f$ (see SI-5) but, in general, it
can be only determined numerically. For the sake of illustration,
results for the function $f$ considered in Eq.(1) are shown in the
Fig.\ref{Jensen} (inset) for the particular case
$\gamma=\gamma_c$. Observe that $F(\tilde{\gamma},s)$ is positive for
$s <1/2$, negative for $s > 1/2$ and vanishes at $s=1/2$ explaining
why the steady state is precisely $s=1/2$ at $\gamma_c$.  Similar
arguments work for other values of $\tilde{\gamma}$. Let us emphasize
that the magnitude of the force decreases as $k$ grows
(Fig.\ref{Jensen}, inset) vanishing in
the limit in which networks are no longer sparse.

Summing up, the sparsity-induced Jensen's force $F$ is responsible for
the emergence of a LAI phase in E/I networks below the mean-field critical point,
$\gamma_c$ as well as for a reduction in the overall level of activity
with respect to the mean-field limit in a region above $\gamma_c$.

Let us emphasize that the annealed-network approximation fits
perfectly well all computational results obtained for quenched
networks, with fixed neighbors and intrinsic structural disorder
(we have computationally verified that,
indeed, the quenched and the annealed versions of the model give
identical results; see SI-5).  The reason for this agreement, lies
in the absence of node-to-node correlations within the LAI phase (see
below), which suggests that the annealed approximation is exact in the
large-network limit. For the sake of completeness, we have
computationally verified that the LAI phase emerges also for other
(non-linear) transfer functions, more random (non hyper-regular)
networks as well as for heterogeneous weight distributions (see SI-5).

\subsection{Phase transitions from and to the LAI phase}
Fig.2 reveals the existence of two phase transitions, one at each side
of the LAI phase. Around the left-most one, at $\gamma_c^{e}$, we
performed standard analyses of avalanches, by introducing a single
seed of activity (one active excitatory node) in an otherwise
quiescent state, and analyzed the statistics of the cascades of
activations it triggers.  For both, avalanche sizes and avalanche
durations, we measured scale-free distributions with the standard
exponents of the unbiased branching process \cite{BP,Serena-BP} (see
SI-6). This is not surprising given the un-structured nature of the
network. Further analyses need to be done in lower dimensional systems
to see if this transition from a quiescent to a noise-induced active
phase shares the critical features of standard quiescent-to-active
phase transitions (known to be in the so-called directed percolation
universality class \cite{Henkel,Marro}) or if novel behavior emerges
owing to noise-induced effects.  On the other hand, the second phase
transition, at $\gamma_c=1/(1-2\alpha$) is a remanent of the original
(discontinuous) mean-field one, and signals a (continuous) transition
between states of low activity to high activity ones. This phase
transition also needs further scrutiny to be fully elucidated.  A
detailed analysis of these phase transitions, as well as of their
possible relevance in connection with the hypothesis that the cerebral
cortex might operate at the edge of a critical point
\cite{BP,Schuster,Haimovici,RMP,LG} is left as an open challenge for
future work (see Discussion).

\subsection{Asynchronous-state features} 

The cortical \emph{asynchronous} state is characterized by a number
of key features (see also Methods) including: {\bf i}) Large variability: the
coefficient of variation, $CV$, defined as the ratio of the standard
deviation to the mean of the interspike intervals (i.e. periods of
un-interrupted silence for a given neuron/node) is large, i.e. $CV \ge 1$
\cite{Softky}.  {\bf ii}) The network-averaged pairwise Pearson's
correlation coefficient $PC$ is very low; actually it decays to $0$
with network size reflecting a lack of synchronization or coherent
behavior \cite{Destexhe2009,Harris2011,Renart}.  {\bf iii}) There is a
(short) time lag between excitation and inhibition (E-I lag) meaning that
an excess in excitation is rapidly compensated by an increase in
inhibitory activity, so that inhibition actively de-correlates neural
populations and the network state remains stable, as theoretically
predicted \cite{Renart, Sompo_Async,Brunel-osc2003,Destexhe_Async} and
experimentally confirmed \cite{Okun2008,Yuste}.

\begin{figure}[h]
\centering
\includegraphics[width=0.9\columnwidth]{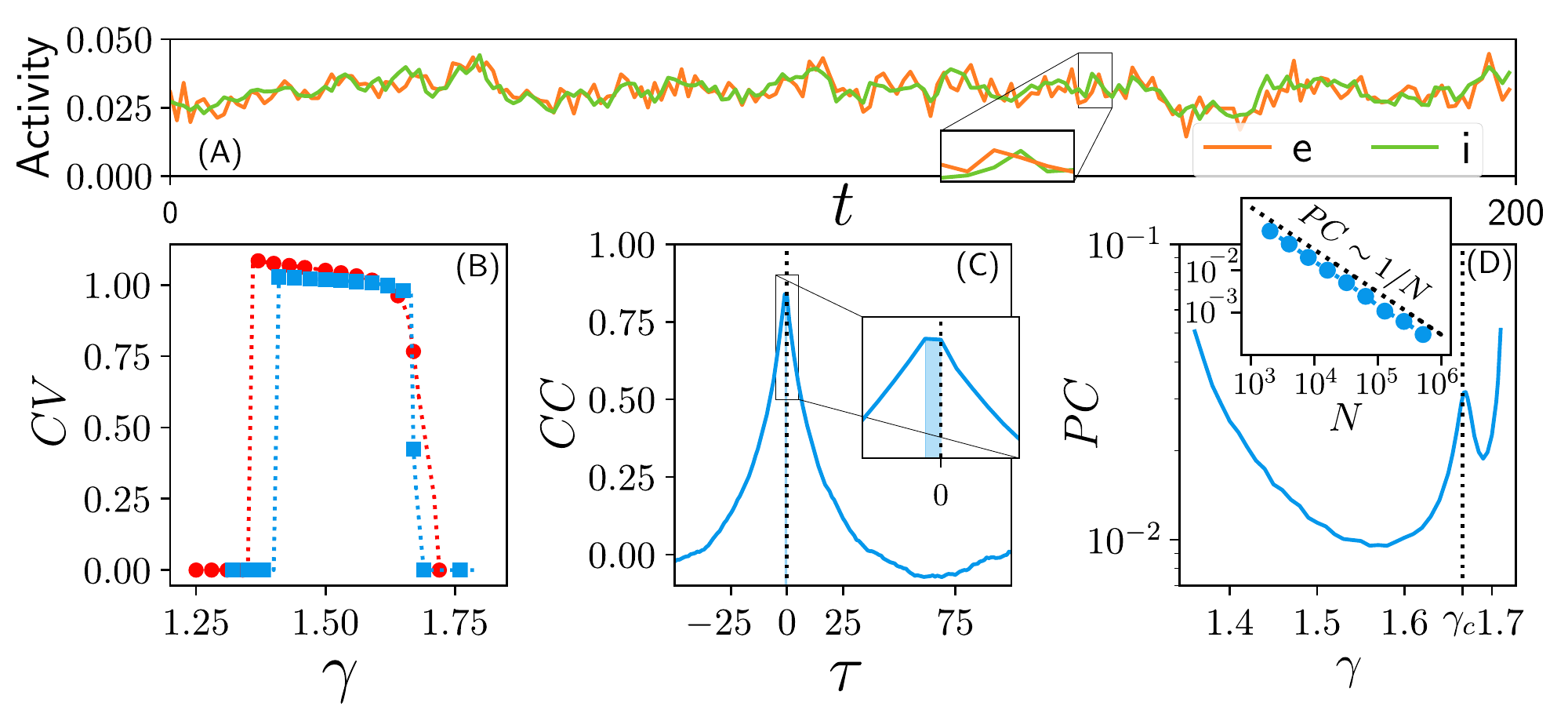}
\caption{(A) Time series of the excitatory (e; orange line) and
  inhibitory (i; green line) network activity in the LAI phase
  (network $N=16.000$). The zoom illustrates the small (one-time step)
  E-I lag present in this phase. (B) Coefficient of variation (CV) vs.
  coupling-strength $\gamma$; $CV \geq 1$ within the LAI phase, while it
  vanishes in the quiescent and active phases (the color code, as in
  Fig.2, stands for connectivity values). (C) Time-lagged
  cross-correlation (CC) between the excitation and inhibition timeseries
  in the LAI phase. The maximum (black dashed line) reflects the
  existence of a one-step E-I lag. (D) Pairwise Pearson's correlation
  (PC) between nodes in the LAI phase as function of $\gamma$; it
  takes small values, but exhibits a marked peak at the critical point
  $\gamma_c$ (dotted line). The inset shows that the PCs scale with
  system size as $1/N$ thus vanishing in the large-network limit (data
  for $\gamma=1.55$, but results valid all across the LAI
  phase).} \label{Correl}
\end{figure}

As shown in Fig.\ref{Correl} the LAI phase --but not the quiescent
nor the active ones-- displays all these key features (see figure
caption for details).
Moreover, in agreement with the original claim for asynchronous states
\cite{vanW-S1,vanW-S2}, we verified that all along the LAI phase (and
only in the LAI phase) the dynamics is chaotic (or quasi-chaotic) in
the sense of damage spreading dynamics \cite{Derrida} (see SI-7). Thus, in synthesis, all the chief features of cortical
asynchronous states are also distinctive and exclusive characteristics
of the LAI phase.

\subsection{Tightly-balanced networks} 
We now scrutinize how the region in parameter space in which the LAI
phase emerges can be maximized, thus limiting the need for parameter
fine tuning to exploit the possible functional advantages of such a
regime. This is achieved by considering \emph{tightly-balanced}
networks (also called detailed-balanced networks)
\cite{Vogels+Abbott,Machens} in which excitatory and inhibitory inputs
are tuned to compensate mutually, so that the average input of
individual nodes is kept close to $0$.  To do so, it suffices to
introduce in the model definition, Eq.(\ref{model}), two different
strengths for excitatory and inhibitory synapses, $\omega^{e}$ and
$\omega^{i}$. In this way, the (leftmost) transition point is shifted
to, $\gamma_{c}^{e}= 1/(\omega^{e}(1-\alpha))$, while $\gamma_c$
changes to
\begin{equation}
\gamma_c = \dfrac{1}{\left(\omega^{e}(1-\alpha)-\omega^{i}\alpha \right)}
\label{balanced}
\end{equation}
which diverges to infinity if
$\omega^{e}/\omega^{i}=\alpha/(1-\alpha)$, implying that the largest
possible LAI phase is obtained when such a condition is met (observe
that in such a limit the level of activity varies very slowly
converging to $s=1/2$ as $\gamma \rightarrow \infty$).  But, given
that $e/(1-\alpha)=i/\alpha$, the above condition corresponds
precisely to the tightly-balanced networks for which the averaged
input of each single node,
$\langle{\Lambda}\rangle =\tilde{\gamma} (\omega^{e} e - \omega^{i}
i)$, vanishes.  Thus, tightly-balanced networks have the largest
possible LAI phase and the largest possible dynamic range.

\subsection{Experimental measurements of the LAI phase and the Jensen's force} 
Is it possible to measure the Jensen's force experimentally? We
believe it is, but explicitly designed setups would be required.
First of all, let us recall that asynchronous states (i.e. LAI phases)
have been detected experimentally both \emph{in vivo} and \emph{in
  vitro} \cite{Trevino,Reyes,McCormick2}.  Importantly, with today's
technology, the spiking activity of more than $1000$ neurons can be
measured simultaneously (see e.g. \cite{Bialek2018}), so that much
better statistics can be collected. In principle, one should be able
to compute the Jensen's force in this type of experiments.  In SI-8 we
propose a tentative experimental protocol to do so. However, we leave
this programme for future research as well as an open challenge for
experimentalists.

\section*{Conclusions and Discussion} 

It has been long observed that neurons in the brain cortex remain
active even in the absence of stimuli \cite{Softky,Arieli,Abeles}.
Often, such a sempiternal spontaneous activity is steady and
highly-irregular --the so-called asynchronous state-- while in some
other circumstances, depending mostly on cortical region and
functional state, diverse levels of synchronization across the
asynchronous-synchronous spectrum are observed
\cite{Latham2000,Brunel1}.

While the role of synchronization in neuronal networks has been long
studied \cite{Buzsaki}, the role of the asynchronous state remained
more elusive \cite{Renart}.  Presently, it has become widely shown
that the asynchronous state emerges from the interplay between
excitation and inhibition, and that it is essential for network
stability and to allow for high computational capabilities
\cite{Yuste,Vogels,recent-Sompo}.

Our main goal here was to investigate the origin low-activity regimes
in excitation/inhibition networks, determining in particular the
nature of their (thermodynamic) phases.  For this, we employed a
statistical-mechanics viewpoint and searched for a modelling approach
as parsimonious as possible, i.e. a sort of Ising model of E/I
networks.  In particular, we analyzed a model which further simplifies
the one proposed by Larremore \emph{et al.}  \cite{Larremore} in a few
different ways.  For example, we removed network heterogeneity both in
its architecture and in the allowed synaptic weights to allow for
mathematical tractability.

Our main result is that E/I networks exhibit a non-trivial LAI phase
in between standard quiescent and active phases, in which activity
reverberates indefinitely without the need of external driving, nor of
intrinsically firing neurons (in contradictions to many previous
beliefs).  Such LAI phase stems purely from fluctuations and, thus, have
little to do with the specific network structure. In particular, this
disproved an initial conjecture of us suggesting that intermediate
levels of activity could be related to so-called \emph{Griffiths
  phases}. Such phases have remarkable features \cite{GPs} and have
been claimed to be relevant for cortical dynamics \cite{Moretti}; they
also emerge in between quiescent and active phases, but only in
systems characterized by structural heterogeneity and, thus, are
unrelated to the novel LAI phase uncovered here. Nevertheless, an
important research line left for future work, is to analyze how the
properties of the LAI phase are altered in more structured and
realistic networks including e.g. broad degree distributions,
clustered structure and modular-hierarchical organization, which might
lead to novel phenomena \cite{Doiron,Doiron2,Sompo-last,Moretti}.

Two key ingredients are necessary for the LAI phase to emerge: a
spontaneously generated dynamical balance between excitation and
inhibition and network sparsity. The resulting phase has all the
statistical properties usually ascribed to asynchronous states.

An issue worth discussing is the dependence of the presented phenomena
on network connectivity and the connection of our work with the
standard view of balanced networks as originally proposed in the
seminal work of van Vreeswijk and Sompolinski\cite{vanW-S1}.  As we
showed, the LAI phase emerges out of input fluctuations and --as the
input standard deviation scales with $\gamma/\sqrt{k}$-- it relies
crucially on the finiteness of $k$, i.e. on network sparsity. However,
it is important to underline that, as we showed, the LAI phase
survives even for arbitrarily large values of $k$, but larger and
larger network sizes $N$ are required for it to be evident.  However,
it is also possible to adopt the original scaling proposed in
\cite{vanW-S1}, where it was argued that if the strength of individual
synapses is of order $1/\sqrt{k}$ (rather than constant as here), it
compensates fluctuations in the number of actual inputs (order
$\sqrt{k}$), leading a total input fluctuations of the same order of
the neuron firing threshold (order unity), and thus to
fluctuation-controlled activations. This type of scaling can be easily
accommodated within our approach just by replacing $\gamma/k$ in Eq.1
by $\gamma'/\sqrt{k}$; with this scaling, the critical points in
terms of the new coupling constant $\gamma'$ are shifted as $k$
grows, and the noise-induced phase persists even in the limit of
non-sparse networks.  Note also that, as illustrated here, having a
sharp threshold is not a necessary ingredient for the phenomenon to
occur: the LAI phase also emerges when considering, e.g. a transfer
function such as the hyperbolic-tangent without a hard
discontinuity. In other words: the Jensen's force is more
general that a hard threshold, noise-filtering, mechanism.

In order to verify whether more realistic neuronal networks models
exhibit also an intermediate phase, in between quiescent and standard
active ones, we first scrutinized the recent literature.  We found
that there are two recent computational analyses of E/I networks of
integrate-and-fire neurons with (current-based or conductance-based)
synapses confirming the emergence of a similar self-sustained
intermediate regime with high variability
\cite{Einevoll,brasileiros}. This confirms that the very general
mechanism put forward here also applies to more detailed/complicated
neuron models. Furthermore, the concept of Jensen's force
sheds light on the computational findings of these recent works.

We have also proposed a tentative protocol to challenge
experimentalist to empirically measure Jensen's forces in
actual neuronal networks, either \emph{in vivo} or \emph{in
  vitro}. Even if technical difficulties are likely to emerge, we
strongly believe that Jensen's forces are susceptible to be observed
and quantified in the lab. This research programme, if completed,
would strongly contribute to shedding light on the noisy dynamics of
cortical networks, as well as on the way it helps processing
information.

Let us also comment on the relationship between the so called
``criticality hypothesis'' --i.e. the idea that the cortex, as well as
some other biological systems, might extract important functional
advantages from operating near the critical point of a continuous
phase transition \cite{BP,Schuster,Mora-Bialek,Shew2011,RMP,LG}-- and
the findings in this work.  Let us emphasize that asynchronous states
and critical states have almost opposite features: the first is
characterized by active de-correlation of nodes and the second
exhibits strong system-spanning correlations. Thus Clarifying the
interplay between these two antagonistic interpretations/phenomena
--and analyzing them together within a unified framework-- is a
challenging goal \cite{Viola2018,Stepp-Plenz}. We believe that our
simple model (probably improved with further important
  ingredients such as some for of synaptic plasticity (as e.g. in
  \cite{LG})) is a good candidate to constitute a unified framework
to put together asynchronous and synchronous states and the critical
phase transition in between, and to analyze these fundamental
questions. Observe in particular that the LAI phase is separated from
the quiescent and active phases, respectively, by continuous phase
transitions --including critical points-- whose specific details still
need to be further elucidated. As a matter of fact, having a good
understanding of the main phase transitions of E/I networks is a
fundamental preliminary step to make solid progress to contribute to
the criticality hypothesis.

Finally, let us mention that we are presently exploring the
possibility of observing similar LAI phases in other biological
networks such as gene regulatory ones, where gene repression plays a
role equivalent to synaptic inhibition in neural networks where
opposite conflicting influences may mutually compensate to each other,
leading to noise-induced phenomena.  We hope that the novel stochastic
force and phase elucidated here foster new research along and this and
similar lines.

\section*{Methods}

Some of the most relevant methods have been sketched in the main
text. Here we detail some important methodological aspects. Further
details are provided in the Supplementary information (SI).

\subsection*{Hyper-regular networks}

For the sake of mathematical tractability, we consider hyper-regular
networks in which each node has exactly $k_{exc}=k(1 - \alpha)$
excitatory neighbors and $k_{inh}=k\alpha$ inhibitory ones pointing to
it (where $\alpha$ is the fraction of inhibitory nodes). For this, we
follow these steps: (i) two random regular networks, one of excitatory
nodes with connectivity $k_{exc}$ and one of inhibitory units with
connectivity $k_{inh}$ are generated; (ii) $k^e=k(1 - \alpha)$ links
(avoiding node repetitions) are randomly chosen to point to each
inhibitory node. This process got sometimes stuck due to a topological
conflict, so we re-started the process after $10^6$ unsuccessful
attempts to include new links.  Each link of the so constructed
networks is taken with positive weight for interactions from a
excitatory nodes $j$ to a inhibitory neuron $i$ ($\omega_{ij} >0$) and
negative for the opposite interaction ($\omega_{ji} < 0$). On the
other hand, all weights with the excitatory (resp. inhibitory)
subnetwork are positive (resp. negative).  For the purpose of
illustration an hyper-regular network is shown in SI-1.

\subsection*{Mean field approach}
The excitatory and inhibitory populations (E,I) evolve stochastically
according to a Master equation \cite{Gardiner} described by the
following transition rates for large networks:
\begin{eqnarray}
  \Omega\left(E, I\rightarrow E+1, I\right)&=
  &\left[N\left(1-\alpha\right)-E\right]\langle
    f\left(\varLambda\right)\rangle \nonumber \\
  \Omega\left(E, I\rightarrow E-1, I\right)&= &E\left[1-\langle
                                                f\left(\varLambda\right)\rangle\right]
  \nonumber\\
  \Omega\left(E, I\rightarrow E, I+1\right)&
                                             =&\left[N\alpha-I\right]\langle f\left(\varLambda\right)\rangle \nonumber \\
  \Omega\left(E, I\rightarrow E, I-1\right)& = & I\left[1-\langle f\left(\varLambda\right)\rangle\right]
\end{eqnarray}
where the timescale has been set to unity, $\langle f(\Lambda)\rangle$
is the average probability for any given node to become active
($\langle\cdot\rangle$ stands for network average), and factors such
as $N\left(1-\alpha\right)-E$ (resp. $(N\alpha-I)$) describe the
number of inactive excitatory (resp. inhibitory) nodes. Performing a
$1/N$ expansion of the corresponding Master equation \cite{Gardiner}
and keeping terms up to leading-order, one readily obtains the
following deterministic equations:
\[
\begin{cases}
\dot{e}=\left(1-\alpha\right)\langle f\left(\Lambda\right)\rangle-e\\
\dot{i}=\alpha\langle f\left(\Lambda\right)\rangle-i
\end{cases}
\]
where the dot stands for time derivative for $e=E/N$ and $i=I/N$,
respectively. In particular, considering a fully-connected system in
the large size limit (i.e. $N\rightarrow\infty$), fluctuations in the
input of each node are negligible. Thus, all nodes receive the same
input, and the mean of the transfer function values is replaced by the
transfer function of the mean input, i.e. the mean-field approach
implies
\begin{equation}
 \langle f(\Lambda)\rangle=f(\langle \Lambda \rangle)
\end{equation}
The detailed procedure to compute these averages is presented in SI-4.

\subsection*{Asynchronous-state features}
\paragraph*{Coefficient of variation} It is defined as the quotient of
the standard deviation $\sigma_{ISI}$ to the mean $\mu_{ISI}$ of the inter-spike interval
($ISI$) on individual units:
\begin{equation}
 CV=\frac{\sigma_{ISI}}{\mu_{ISI}}.
\end{equation}

\paragraph*{Excitatory/inhibitory cross-correlation} Given two time series $x(t)$ and
$y(t)$, the Pearson correlation coefficient of $x(t)$ and
$y(t+\tau)$
\begin{equation}
CC(\tau) = \frac{1}{\sigma_x \sigma_y}\sum_{\tau=-\infty} ^{+\infty} \overline{x(t)} y(t+\tau).
\end{equation}
where $\sigma_x$ and $\sigma_y$ are the standard deviations of the
time series $x(t)$ and $y(t)$, respectively and $\tau$ is a time
delay.  Since we are interested in the E/I lag, we substract the mean
from the time series, i.e. we take $x(t)=e(t)-\mu_e$ and
$y(t)=i(t)-\mu_i$. This procedure ensure us a correct normalization,
so $CC(\tau)\in[-1,1]$. In this way, if $CC(\tau)$ has a peak for
$\tau<0$, we conclude that the activity of the inhibitory population
resembles the activity of the excitatory one, but it is shifted to the
left: excitatory population spikes first and it is followed by the
inhibitory one.

\paragraph*{Pairwise correlation} The Pearson's correlation
coefficient between a randomly selected pair of nodes in the network,
$x_i$ and $x_j$, is defined as
\begin{equation}
 PC_{x_i,x_j}={\frac {\langle x_ix_j \rangle - \langle x_i\rangle \langle x_j \rangle}{{\sqrt {\langle x_i^{2}\rangle-\langle x_i\rangle ^{2}}}~{\sqrt {\langle x_j^{2}\rangle-\langle x_j\rangle ^{2}}}}}
\end{equation}
where $\langle \cdot \rangle$ represents a temporal average. The total
Pearson's correlation coefficient ($PC$) is computed by averaging
$PC_{x_i,x_j}$ over $500$ pairs of nodes for different realizations.

\paragraph*{Chaotic behavior} It is important to scrutinize the
possible chaotic nature of the LAI phase \cite{vanW-S1}. For this, we
employ the standard method (particular results are shown in SI-7),
consisting in analyzing the dynamics of damage spreading
\cite{Derrida}. The method involves the next steps: (1) take a
specific state of a network, $M$, and a construct an identical replica
of it, $M'$, in which the state of only a randomly-chosen node is
changed; (2) the Hamming distance, $H$ --defined as the difference of
states between $M$ and $M'$-- is commuted after one time step (i.e. an
update of all the nodes of the two networks) and finally, (3) $H$ is
averaged over many realizations (i.e. over different locations of the
initial damage and stochastic trajectories) obtaining the branching
parameter, $B$. If $B<1$ perturbations tend to shrink and the network
is in a ordered phase, while if $B>1$ perturbations growth on average
and the network exhibit chaotic-like behavior. Finally, for marginal
propagation of perturbations, i.e. $B=1$, the network is critical.

\section*{Acknowledgments}
We acknowledge the Spanish Ministry of Science as well as the Agencia
Espa{\~n}ola de Investigaci{\'o}n (AEI) for financial support under
grant FIS2017-84256-P (FEDER funds). V.B and R.B. acknowledge funding
from the INFN BIOPHYS project. We warmly thank P. Moretti, J. Mejias,
for very useful comments.  This work is dedicated to the memory of
Prof. Daniel Amit, who shared with us (MAM), his pasion for
understanding cortical fluctuations before this was a popular
research topic.

\newpage

\thispagestyle{empty}


\section*{Supplementary material}

# 1 Hyper-regular networks

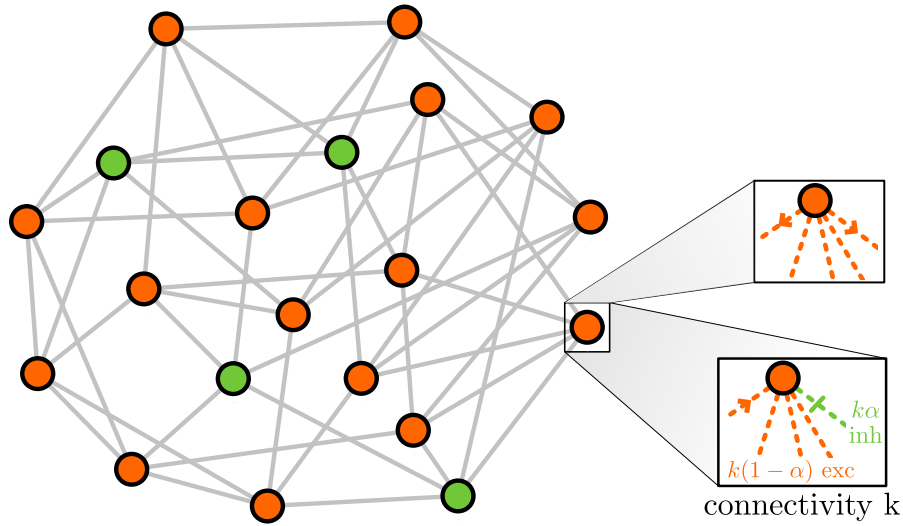

Figure S1: Hyper-regular network with $N = 20$ nodes and connectivity $k = 5$. As in the main text, orange nodes stand for excitation and green nodes for inhibition. For the zoomed node, the difference between out-activity and in-activity is also shown (i.e. each node has $k = 5$ excitatory (or inhibitory) outbound links as well as $k(1-\alpha)$ excitatory and $k\alpha$ inhibitory inbound links). In particular, each node has, thus, 5 inbound inputs of which 4 are excitatory and 1 inhibitory, as well as 5 outbound links: all of them positive por excitatory units and negative for inhibitory ones.

# 2 Mean-field phase diagram for non-linear transfer functions

In the mean-field approximation, $\langle f(\Lambda) \rangle \simeq f(\langle \Lambda \rangle) = f(\gamma(1-2\alpha)s)$ (see derivation below). Then, the equation for the activity is simply

$$\dot{s} = f(\gamma(1-2\alpha)s) - s, \qquad (1)$$

which has fixed points $f(\gamma(1-2\alpha)s^*) = s^*$. While the position of the critical point $\gamma_c$ is robust under changes of the transfer function, the stability of the fixed points may change. As a consequence, the shape of the mean-field phase diagram strongly depends on the choice of the transfer function. Since, in mean-field, the activity is bounded to lie in the interval $0 \leq \Lambda \leq 1$, it is possible to Taylor expand $f(\Lambda)$ around $\Lambda = 0$ (which is always a fixed point):

$$f(\Lambda) = a\Lambda + b\Lambda^2 + c\Lambda^3 + \ldots \qquad (2)$$

where $a, b$ and $c$ are un-specified parameters. This is often called a *Landau expansion* in the theory of critical phenomena [1]. Note that the linear term in Eq.(1) can be absorbed into the $a$ term of Eq.(2), just shifting its value. The signs of the coefficients $a$, $b$, $c$... determine the behaviour of stable fixed points of the system. Fig.S2 shows the phase transition arising from different combinations of these coefficients. In particular, the rightmost figure shows the most realistic case in which there is a region of bistability with hysteresis.

Some of the key properties of these phase transitions persist even for sparse networks, where the LAI phase appears, as shown in Appendix 5.

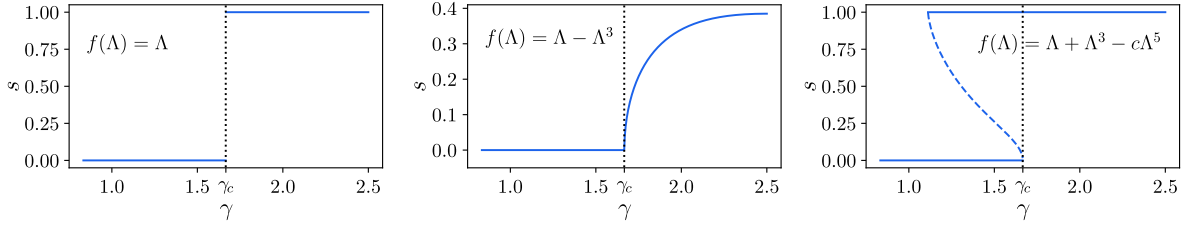

Figure S2: Mean-field phase diagram of Eq.(1) for different choices of the transfer function $f(\langle \Lambda \rangle) = f(\gamma(1-2\alpha)s)$. Depending on the signs of the coefficients chosen in the Landau expansion, we may observe discontinuous, continuous, or discontinuous with bistability types of bifurcations (the discontinuous blue line represents unstable fixed points).

## 3 Distribution of the input of an individual neuron for annealed networks

Here we discuss in detail the derivation of the distribution of inputs received by each individual node within the annealed version of the model. If node has $j$ excitatory active neighbors and $l$ inhibitory active ones, its input is given by $\Lambda_{jl} = \tilde{\gamma}(j-l)$, with $\tilde{\gamma} = \gamma/k$. At each timestep any given node chooses randomly $k$ connecting nodes: $n$ inhibitory and $k-n$ excitatory ones. Thus, the probability to have an input of value $\Lambda_{jl}$ is controlled by the probability distribution of having $j$ excitatory and $l$ inhibitory active neighbors $p(\Lambda_{jl}|n) = p(j|k-n)p(l|n)$. If the average activity of the network is $s$, a node picked randomly will be active with probability $s$, regardless of whether it is excitatory or inhibitory. As a consequence, the probability of finding $l$ active inhibitory neighbors out of the total $n$ inhibitory connections is given by the binomial distribution:

$$p(l|n) = \binom{n}{l} s^l (1-s)^{n-l}. \tag{3}$$

Similarly, the probability $p(j|n)$ to have a $j$ excitatory active nodes out of $k-n$ possible ones is another binomial distribution. In the case of the hyper-regular model, in which we the number of inhibitory neighbors can be written as $n = k\alpha$ one readily obtains

$$p(\Lambda_{jl}) = \sum_n \delta_{n,k\alpha} p(\Lambda_{jl}|n) =$$
$$= \binom{k\alpha}{l}\binom{k(1-\alpha)}{j} s^{j+l}(1-s)^{k-j-l}, \tag{4}$$

(while for regular but not hyper-regular networks this expresion becomes more involved). Note that the above probability depends exclusively on the average activity of the network, $s$, and the connectivity $k$, so $p(\Lambda_{jl}) \equiv p_{lj}(s)$, as defined in the main text. From this, it is possible to evaluate the averages of any function of the input. In particular, the first and second moments of the input can be easily computed

$$\langle \Lambda \rangle = \sum_{j=0}^{k-n}\sum_{l=0}^{n} p_{lj}(s)\Lambda_{jl} = \tilde{\gamma}ks(1-2\alpha) = \gamma s(1-2\alpha)$$
$$\langle \Lambda^2 \rangle = \tilde{\gamma}^2 \left[ks^2\left((1-2\alpha)^2 k - 1\right) + ks\right],$$

and, similarly, for the variance

$$\sigma^2(\Lambda) = \tilde{\gamma}^2 ks(1-s) = \gamma^2 s(1-s)/k \tag{5}$$

3

This implies that fluctuations of the input –as measured by the standard deviation $\sigma_s$– are proportional to $1/\sqrt{k}$, as expected from the central limit theorem [2, 3]. Also, it follows that the states ($s = 0$ and $s = 1$) exhibit no fluctuations, and that the maximum level of fluctuations occurs at $s = 1/2$, which coincides with the value of the activity at the critical point $\gamma_c$.

Moreover, we assumed that each node has $k$ inbound connections and a fixed number $n$ of inhibitory neighbors, but, as a matter of fact, it is also possible to consider the non-regular network case by letting $k$ to be a random variable itself, distributed according to some arbitrary probability distribution $g(k)$. This generalization changes Eq.(4) to

$$p(\Lambda_{jl}) = \sum_{k=1}^{+\infty} \sum_{n=0}^{k} \binom{n}{l} \binom{k-n}{j} g(k) \cdot$$
$$\cdot h(n|k) s^{j+l} (1-s)^{k-j-l} \qquad (6)$$

where $h(n|k)$ is the probability of having $n$ inhibitory neighbors, given a connectivity $k$. Although the sum cannot be worked out analytically, it is still possible to work it out numerically. In particular, letting $g(k)$ to be a Poisson distribution and $h(n|k)$ a binomial distribution with probability $\alpha$, it is possible to obtain results for Erdős-Rényi networks.

## 4   Derivation of the critical and saturation points

Considering the distribution of possible inputs for each node in the annealed approximation, it is straightforward to derive the critical and saturation points of the system. The equation for the activity reads $\dot{s} = \langle f(\Lambda) \rangle - s$. Given Eq.(4) it is possible to compute exactly the average value $\langle f(\Lambda) \rangle$:

$$\langle f(\Lambda) \rangle = \sum_{j=0}^{k-n} \sum_{l=0}^{n} f(\tilde{\gamma}(j-l)) p_{lj}(s), \qquad (7)$$

where $n$ is the number of inhibitory neighbors of a node, and $j$ and $l$ represent the number of active excitatory and inhibitory neighbors, respectively. Although for the hyper-regular case $n = k\alpha$, we consider here a generic $n$ value and replace it by its value at the end of the calculation. Due to the non-linearity of the function $f$, it is not possible to fully solve the problem analytically. However, it is still possible to derive relevant information from equation (7). First of all, near the critical point $\gamma_c^e$, one expects to have a very low level of activity, which suggests to expand Eq.(7) up to first order in $s$. The probability $p_{lj}(s)$ contains a factor $s^{j+l}(1-s)^{k-j-l}$, so that, if $k$ is large enough (and taking into account that for $\alpha = 0.2$ we should assume $k > 5$ here), all terms with $j + l \geq 2$ contribute to second order in $s$. Thus, the $(j, l)$ pairs that contribute to first order are just $(0, 0)$, $(0, 1)$ and $(1, 0)$. Of these terms, note that $f(\Lambda_{00}) = f(\Lambda_{01}) = 0$, so, the only contributing one is $(1, 0)$. Taking this value in Eq.(7) and performing the Taylor expansion, one readily obtains

$$\langle f(\Lambda) \rangle = f(\tilde{\gamma}) \binom{n}{0} \binom{k-n}{1} s (1-s)^{k-1} \simeq$$
$$\simeq f(\tilde{\gamma}) k (1-\alpha) s. \qquad (8)$$

Considering also that $f(\tilde{\gamma}) = \tilde{\gamma}$, we obtain

$$\dot{s} = \gamma(1-\alpha)s - s, \qquad (9)$$

that exhibits a bifurcation at $\gamma_c^e = 1/(1-\alpha)$ separating the quiescent phase ($s = 0$) from the LAI phase ($s \neq 0$).

In order to obtain information about the saturation of the activity one can use the very same procedure to expand around $s = 1$. Observe that now the three pairs that contribute to

first order in $(1-s)$ are $(k-n, n)$, $(k-n-1, n)$ and $(k-n, n-1)$, and $f(\Lambda_{jl})$ does not vanish for any of these terms. Introduce such values in Eq.(7), and after Taylor expanding, one finds:

$$\langle f(\Lambda) \rangle = f\left[\tilde{\gamma}(k-2n)\right]\left[1 + k(1-s)\right] + \\ + (1-s)(k-n)f\left[\tilde{\gamma}(k-2n-1)\right] + \\ + nf\left[\tilde{\gamma}(k-2n+1)\right]. \tag{10}$$

To short the notation, we define

$$f(\Lambda_0) \equiv f\left[\tilde{\gamma}(k-2n)\right]\left[1 + k(1-s)\right], \\ f(\Lambda_-) \equiv f\left[\tilde{\gamma}(k-2n-1)\right], \\ f(\Lambda_+) \equiv f\left[\tilde{\gamma}(k-2n+1)\right],$$

where all $\Lambda$ implicitly depend on $\gamma$. Thus, depending on the value of the synaptic strength, some the above terms can saturate, i.e. $f(\Lambda) = 1$. In fact, if we that $\gamma > \gamma_c$, then $f(\Lambda_+) = f(\Lambda_0) = 1$. Under this assumption, replacing the value at which the bifurcation happens in Eq.(1), we obtain

$$\gamma^{sat} = \frac{1 - k(1-\alpha)}{(1-\alpha) - k(1-\alpha)(1-2\alpha)}, \tag{11}$$

which coincides very accurately with the point at which the activity becomes $s = 1$ in the sparse model, as observed in numerical simulations (see Fig.S3). Note that at the limit $k \to +\infty$ this coincides with the mean-field critical point, but for sparse networks, saturation occurs after $\gamma_c$. Note that the mean-field can be recovered again if we let all the three functions to be $f(\Lambda) = \Lambda$, i.e. none of them saturates.

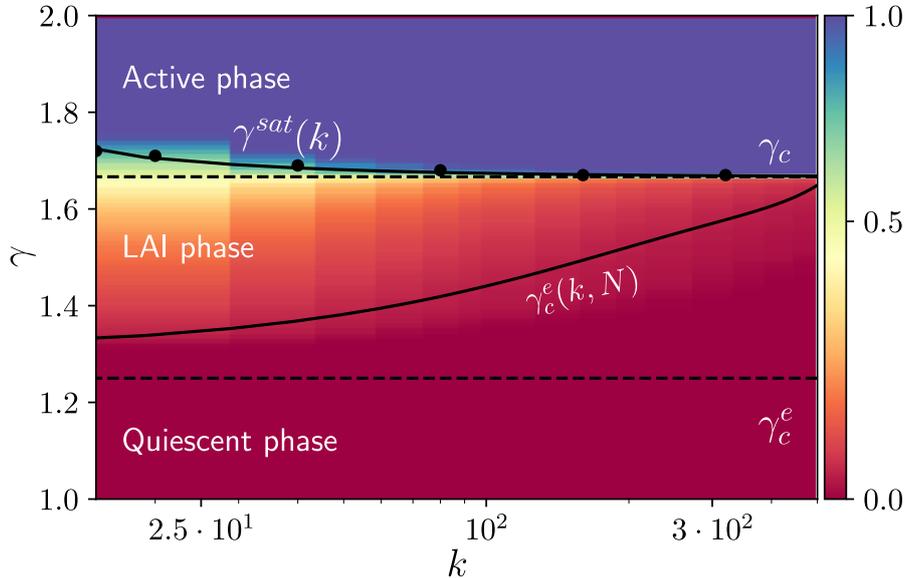

Figure S3: Phase diagram as a function coupling-strength ($\gamma$) and connectivity $k$ for a finite size $N = 16000$ nodes; color code indicates the level of averaged overall activity $s$. This shifts from the quiescent phase (reddish colors) to the active phase (blueish colors). Horizontal dashed lines correspond to the critical points $\gamma_c^e$ and $\gamma_c$ at the large-N (thermodynamic) limit. The saturation value $\gamma^{sat}(k)$ corresponds to Eq.(11); results from simulation are marked as black points. The curve $\gamma_c^e(k, N)$ represents an interpolation of the values obtained from simulations, coincides with the dashed line in the large-N limit.

# 5 Robustness of the results

Here we explore the robustness of the results and conclusions presented in the main text with respect to the modification of various modelling choices.

**Changes in the transfer function**

First we discuss modifications in the transfer function $f(\Lambda)$, that represents the probability of activation of a neuron with input $\Lambda$. The characteristics of the mean-field phase transition, depend on the particular election of this function, as discussed in Appendix 2: it can be either continuous, discontinuous, or discontinuous with bistability and hysteresis. For example, if we use $f(\Lambda) = \tanh(\Lambda)$ for $s > 0$ (and $f(s) = 0$ otherwise), the order of the transition changes from first order to second order. We have computationally verified that the presence of the self-sustained LAI phase is not altered by the choice of $f(\Lambda)$; it exists both in the case in which the mean-field transition is continuous (see Fig.S4) and discontinuous with bistability and hysteresis (see Fig.S5).

In particular, we wondered whether the asynchronous irregular spiking and other characteristic features of the LAI phase –as reported in the main text– are exclusive of the LAI phase or also appear in the active phase in the case in which this exhibits a low level of activity (e.g. in the case of a continuous phase transition, such as that of Fig.6 central). Fig.S4) shows the coefficient of variation ($CV$) and time-lagged cross-correlation ($CC$), revealing that both of these quantities are approximately constant and large inside the LAI phase, but they decay quickly as soon as the coupling parameter $\gamma$ enters the active phase. We conclude that the dynamical properties of the characteristic features of the LAI phase cannot be found in a regular active phase, even when this also displays relatively low activity.

In the case of the continuous mean-field transition, we can take advantage of the functional form of the hyperbolic tangent, in order to compute analytically an expression for the Jensen's force. If the response function $f(\Lambda)$ can be expanded in Taylor series (for $s > 0$), we can approximate the Jensen's stochastic force as

$$F(\Lambda) = \sum_{n=0}^{+\infty} \frac{f^{(n)}(0)}{n!} \left( \langle \Lambda^n \rangle - \langle \Lambda \rangle^n \right), \qquad (12)$$

for positive inputs. The moments of the input distribution may be computed as shown in the previous Appendix, and thus it is possible to obtain an analytic approximation to the desired order $n$. If we use $f(\Lambda \geq 0) = \tanh \Lambda$, performing the expansion[1] we obtain that the first term that contributes to the Jensen's force is the one that corresponds to the third moment,

$$F(\Lambda) \equiv F(s) \simeq \frac{\gamma^3}{3} \alpha(1-\alpha)(1-2\alpha)ks(1-s)(1-2s). \qquad (13)$$

Observe that in this case the Jensen's force vanishes at $s = 0, 1/2, 1$, in agreement with what we noticed numerically in the main text for the case of the linear piecewise function. In particular, let us also remark that –in the detailed-balanced case– where the mean-field term vanishes and the dynamics is given $\dot{s} = F(s)$, $F(s)$ is of the form [2] $F(s) \sim as - bs^2$, showing the continuous quiescent-active transition.

---

[1] Since we use $f(\Lambda < 0) = 0$, the response function is not analytic at the origin –hence, it cannot be Taylor expanded at 0. However, we assume that the expansion is formally valid for all $\Lambda > 0$. A similar calculation cannot be performed for the linear piecewise function, due to the saturation condition $f(\Lambda \geq 1) = 1$.

[2] Since $0 \leq s \leq 1$, the $+cs^3$ does not destabilize the system, and can be ignored.

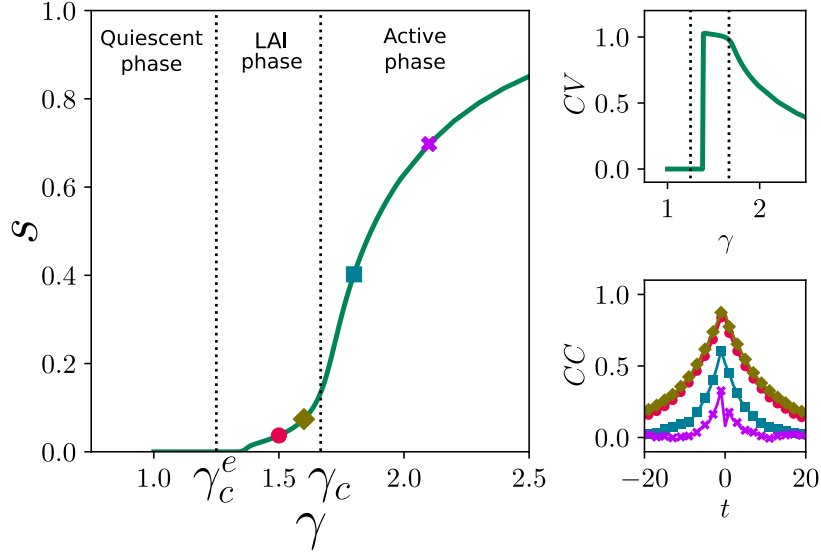

Figure S4: (a) Phase diagram for a model using a hyperbolic tangent (for $s > 0$ and 0 otherwise) as a transfer function in a hyper-regular network with $N = 16000$ nodes and $k = 40$. The critical points, $\gamma_c^e$ and $\gamma_c$, are marked with dotted lines. Observe the presence of an intermediate (LAI) phase, as well as an active phase that emerges after a second-order, continuous phase transition. (b) Coefficient of variation as a function of the coupling $\gamma$. Note that $CV$ is approximately constant inside the LAI phase, where it takes a large value. The value of the $CV$, however, decreases as soon as one enters the active phase. (c) Time-lagged cross-correlations $CC$ for the points marked with symbols in the panel (a). Again, as one goes inside the active phase, the maximum of the correlation starts decreasing, meaning that excitation and inhibition become progressively decorrelated within the active phase.

In the case in which the mean-field phase transition is discontinuous with bistability and hysteresis (see Appendix 2), it is also pertinent to ask whether the emerging LAI phase is able to coexist with the active phase. In order to do that, we consider $f(\Lambda) = \Lambda + \Lambda^3 - c\Lambda^5$ and run simulations for different initial activities $s_0$ [3]. Results are presented in Fig.S5: there is coexistence between the LAI phase (with low but non-vanishing activity) and the active phase. Low initial activity values end up flowing to the LAI state, while higher values of $s_0$ drive the system to the active phase, with larger values of the activity, illustrating the bistability of the dynamics. When $\gamma$ is increased, $s$ slowly increases, until the activity value goes over the instability point (smaller than the mean-field value $\gamma_c = 1/(1-2\alpha) = 1.667$). Above this point, the LAI phase becomes unstable and the system jumps into the active phase. Thus the system exhibits bistability and hysteresis, between the LAI phase and the regular active one.

---

[3]This transfer function is similar to the truncated Taylor expansion of the hyperbolic tangent, changing signs of the non-linear term coefficients. The $c$ term controls the width of the hysteresis cycle, and has been set to $c = 1/2$.

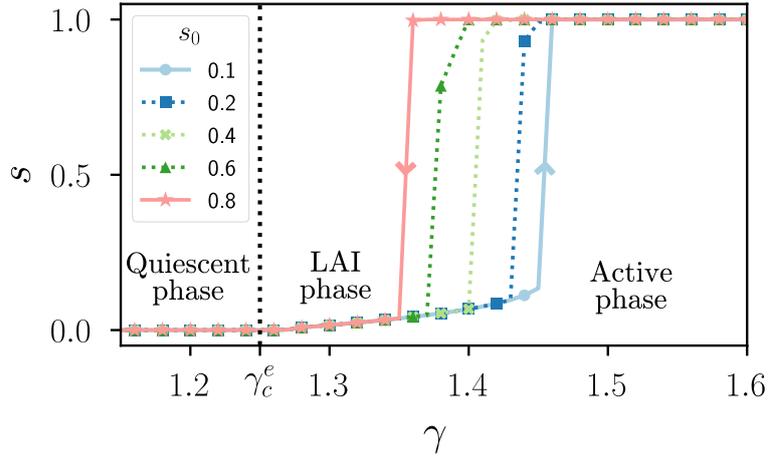

Figure S5: Phase diagram for a transfer function $f(\Lambda) = \Lambda + \Lambda^3 - \frac{1}{2}\Lambda^5$, for different initial states $s_0$ (see legend). The system displays bistability between the LAI phase and the active phase in a full interval of $\gamma$ values. $N = 128000$ and $k = 15$.

### Changes in the network structure

Simulations for annealed and quenched hyper-regular networks show that both cases give the same results computationally, and that such results coincide with the analytic predictions. Thus, as analytical computations are exact only in the annealed case, the essence of the involved mechanism has nothing to do with the network specific structure and, thus, spectral analyses of the connectivity matrix [4, 1] do not add much to the understanding of the noise-induced intermediate phase.

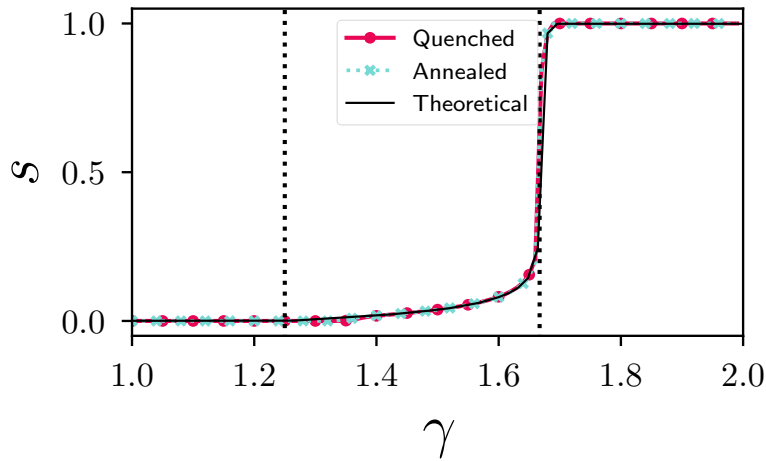

Figure S6: Comparison between annealed and quenched networks (simulation results marked with symbols) and the theoretical prediction for hyper-regular networks with connectivity $k = 40$ ad size $N = 16000$. The agreement between quenched and annealed hyper-regular networks is excellent (perfect within numerical accuracy), and the agreement between these two and analytical predictions is also excellent for sufficiently large network sizes.

We verified computationally that the main results also emerge in more irregular networks. In particular, simulations on Erdős-Rényi networks also reveal the emergence of a LAI phase, as shown in Fig.S7. Similarly, considering a Gaussian distribution of weights with variance $\sigma^2 = 1$ (rather than $\pm 1$), does not affect either the existence of a LAI phase (see Fig.S7).

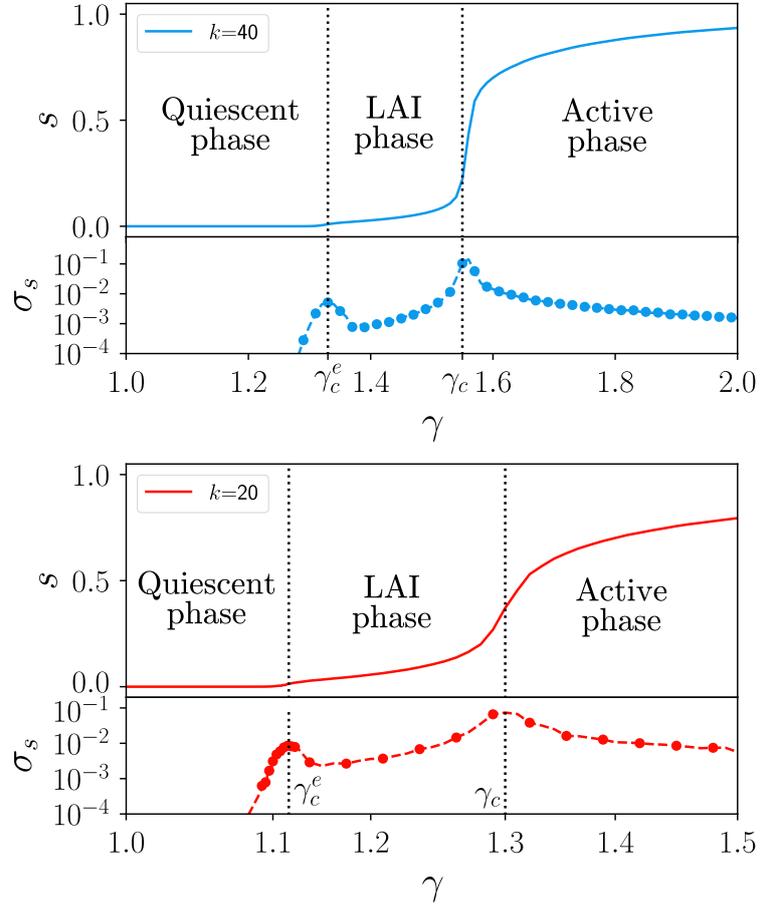

Figure S7: (Up) The LAI phase emerges also in (non-regular) unweighted Erdős-Rényi networks with mean connectivity $k = 40$ and $N = 16000$ as well as in (Down) Erdős-Rényi networks with a Gaussian weight distribution, $k = 20$ and $N = 16000$.

# 6 Avalanches at criticality

Here we scrutinize the dynamics at the noise-induced critical point, separating the quiescent from the LAI phase. In particular, we analyze the emergence of avalanches of activity originated after introducing a seed of activity into an otherwise quiescent state. We observed computationally that at the quiescent-active critical point $\gamma_c^e$ the system displays avalanches –whose sizes and durations are distributed as power-laws as $P(S) \sim S^{-\tau}$ and $P(T) \sim T^{-\alpha}$, respectively– thus compatible with those of the unbiased branching process (see Fig.S8).

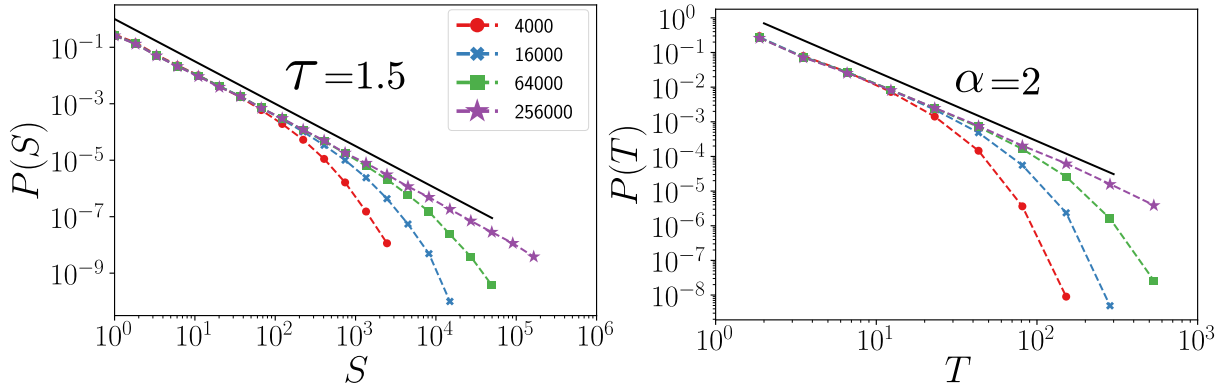

Figure S8: Distribution of avalanche sizes (left) and durations (right) at the critical point $\gamma_c^e$ for different system sizes (see legend) in a hyper-regular network with $k = 15$. Black dotted lines are guides to the eye showing the theoretical values for the unbiased branching process.

On the other hand, at the second critical point, $\gamma_c$ –separating the LAI from the active phase– there are not scale-free avalanches, but just excursions of the global activity around its mean value (results not shown).

# 7 Damage spreading in the asynchronous state

As shown in Fig.S9a, all across the LAI phase we observe a value of $B > 1$, and, as a consequence, chaotic behavior, as previously suggested for asynchronous states [2]. Moreover, as shown in Fig.S9b, flipping a small number of nodes (e.g. 10 nodes) the networks fulfill completely different dynamical states in the LAI phase. Thus, by computing the difference between states in $M$ and $M'$, averaged in time for sufficiently large times ($H_{st}$), one observes that –within the LAI phase– this difference takes almost the same value as the network activity, revealing that the active sites become rapidly uncorrelated in both replicas, reflecting again the chaotic nature of the LAI phase.

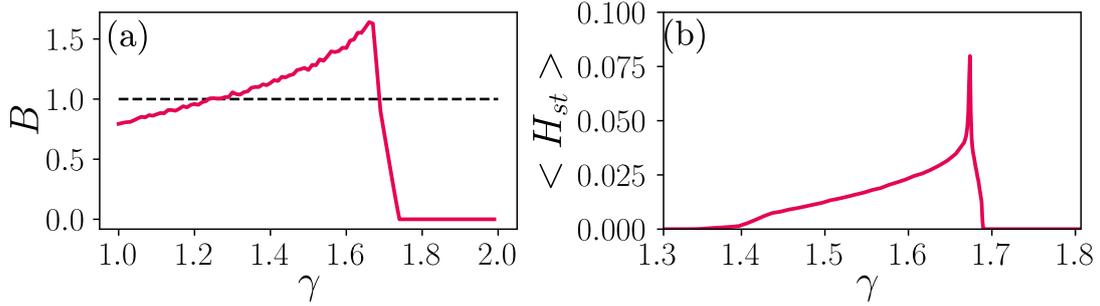

Figure S9: (a) Branching function $B$ in damage spreading experiments (averaged over $10^4$ runs). Black dotted lines represent marginal propagation of activity, i.e. critical dynamics. All across the LAI phase, the dynamics propagates in a chaotic way $B > 1$, while in the quiescent and active phases, the Hamming distance is smaller than 1. (b) Average over runs for the time-averaged Hamming distance in the steady state $\langle H_{st} \rangle$, over $T = 10^4$ MonteCarlo steps; the two initial replicas are different in a small number (10) of nodes. In this case, within the LAI phase the difference between the two replicas $\langle H_{st} \rangle$ is very close to the steady state density, indicating that activity becomes uncorrelated between them (node states coincide only by chance). Simulations run for hyper-regular networks with $N = 16000$, $k = 40$ and $\alpha = 0.2$.

## 8 Experimental measurement of Jensen's force

In order to explicitly observe and measure Jensen's forces in the laboratory, we propose the following (preliminary) experimental protocol:

(i) Consider observations of neuronal activity in the asynchronous state such as those already in the existing literature. Measure the average network activity $s(t)$ as a function of time (in discrete time bins) and determine the value $s(t+1)$ as a function of the activity at the preceding timestep $s = s(t)$ for all possible observed values of $s$ and average across the steady-state time series to obtain good statistics. This procedure provides us with an empirical estimation of the averaged response transfer function as $s(t+1) = \langle f(\Lambda) \rangle_{exp}$ (as easily derived from Eq.(??)). Let us remark that there exist publicly available empirical datasets (e.g. for the rat and cat cortex; see [5] and refs. therein) that can be used for this purpose, though better statistics including many more neurons and longer observations times would be highly desirable.

(ii) Extract individual neurons from the same tissue under study and determine empirically *in vitro* their associated transfer function $f_{exp}(\Lambda)$ (where $\Lambda$ is the input). We believe that this is experimentally feasible as similar measurements have been already successfully performed [6, 7]. If, in the experiments, the responses of diverse neurons are different, a sort of averaged neuron response should be constructed as a proxy for $f_{exp}(\Lambda)$ [6].

(iii) Measuring the membrane potential, it should be possible to estimate the averaged input received by a single neuron in the network, $\langle \Lambda \rangle$ and using the result of (ii) it should be possible to compute $f_{exp}(\langle \Lambda \rangle)$.

(iv) The state-dependent Jensen's force is then determined using Eq.(??), i.e.

$$F(\Lambda) = \langle f(\Lambda) \rangle_{exp} - f_{exp}(\langle \Lambda \rangle). \qquad (14)$$

The theory presented here predicts a non-linear and non-monotonic behavior for $F(\Lambda)$, similar to that of the inset of Fig.3. Note also that the theoretical prediction could be refined and made more specific by implementing in the theoretical model the empirically determined transfer function, $f_{exp}(\Lambda)$. This change might hinder analytical calculations, but would be straightforward to implement in computational analyses of our simple model, that would lead to a specific prediction for the Jensen's force in the experimental setup. In any case, for states

of low activity the difference between the averaged response within the network $\langle f_{exp}(\Lambda) \rangle$ is expected –according to the theory developed here– to be larger than the response of individual neurons to the average activity $f_{exp}(\langle \Lambda \rangle)$, i.e. there should be a repulsive stochastic force, inducing fluctuating states of low-activity, and this should be observed in the experiments.

We leave this programme –which is very likely to need refinements to account for a number of potential experimental pitfalls, such as statistical error from subsampling, time-binning ambiguities, heterogeneous response of individual neurons, etc.– for future research and as an open challenge for experimentalists.



\end{document}